\documentstyle[12pt]{article}
\begin{document}
\thispagestyle{empty}
\begin{center}{\Large{Magnetogenesis from axion and dilaton electrodynamics in torsioned spacetime}}
\end{center}
\vspace{1.0cm}
\begin{center}
{\large By L.C. Garcia de Andrade\footnote{Departamento de
F\'{\i}sica Te\'{o}rica - IF - UERJ - Rua S\~{a}o Francisco Xavier
524, Rio de Janeiro, RJ, Maracan\~{a}, CEP:20550.
e-mail:garcia@dft.if.uerj.br}}
\end{center}
\begin{abstract}
Recently much controversy has been shed on BICEP 2 experiments for the concerning this validity or not and a possible set of new experiments to detect primordial inflation and gravitational waves. Since gravitational waves imply the existence of primordial magnetic fields in this context, C Bonvin, R Durrer and R Marteens [Phys Rev Lett (2014)] have tried to associate the presence of primordial magnetic fields to BICEP 2 by making use of CMB tensor modes. Here we show that by considering torsion dilatonic lagrangean one obtains cosmological magnetic fields of the order of $B\sim{10^{-10}G}$ which may seed galactic dynamos. Actually this new result came out of a mistake of a recent paper published by myself in JCAP (2014). These results are more in accordance with Bamba results [JCAP (2014)] in the context of teleparallel theory of gravity with Einstein's distant parallelism and torsion. These results also support Einstein-Cartan sort of theories of gravity from well-known recent data. Another example which supports the use of modified gravities with torsion to investigate magnetogenesis is the alternative example of using axion electromagnetism with transmutation of torsion into axions to obtain cosmic magnetic seed bound of $10^{-12}G$ this lower bound coincides with Barrow et al [Phys Rev D (2012)] lowest bound of the interval $10^{-20}G$ to $10^{-12}G$ for cosmic magnetic fields in Friedmann universes. \end{abstract}

Key-words: Torsion theories, primordial magnetic fields, galactic
dynamo seeds.
\newpage
\section{Introduction}

Earlier Barrow and Tsagas \cite{1} have shown that in the context of
GR, FRW universes magnetic field can experience superadiabatic
growth with B-field of the order of $B\sim{a^{-1}}$ instead of
$B\sim{a^{-2}}$ of the adiabatic growth. In their case the Maxwell
equations is used instead of more general Maxwell s equations used
to find galactic dynamo seeds \cite{2}. Coupling of vector fields to
spacetime geometry slows down the decay of large scale magnetic
field in open universe, compared to that seen in perfectly flat
models. This results in a large gain in B-field strength during the
pre-galactic era that leads to astrophysically interesting B-fields.
Seeds around $10^{-34}Gauss$ can sustain galactic dynamos if the FRW
universe is dark-energy dominated. It is well-known that dynamo
mechanism requires magnetic seeds with a collapse coherence length
of $100 pc$. This corresponds to a comoving (pre-collapse) scale of
approximately $10 Kpc$. Here we consider seed fields on
curvature-torsion scales of $10 Kpc$, which from dynamo equation
with torsion \cite{3}, yields a cosmological magnetic field of
$10^{-26}Gauss$ in contrast to the $10^{-22}Gauss$ obtained from
cosmic strings \cite{4}. If one still uses the constant torsion, and
the scale of $10Kpc$ the galactic dynamo seed is given by
$B\sim{10^{-10}Gauss}$ which is only one order of magnitude less
than the one obtained for the low limit of cosmic magnetic field
obtained by K. Pandey et al \cite{5} for primordial magnetic field
lines using $Ly{\alpha}$ clouds on $1 Kpc$ scales. To further test
the model we compute the cosmic magnetic fields in comoving cosmic
plasmas with torsion and obtain primordial magnetic seeds of the
order of $B\sim{10^{-25}Gauss}$ which though is weak is still enough
to seed galactic dynamos \cite{6}. Bamba et al \cite{7} has recently show that however when one uses  a teleparallel torsion theory on a dilaton electromagnetismone obtains a value for cosmic magnetic field which is $B\sim{10^{-9}G}$ which shows that there is no need of dynamo amplification. Here by correcting a recent mistake in JCAP paper \cite{8} one obtains results which range between $10^{-5}G$ and $10^{-10}G$
which also does not need the dynamo mechanism for interestellar medium as confirmed in other theories of gravity. As shown recently by Urban \cite{9}, large-scale magnetic fields
could be obtained from QCD axionic dark matter in Minkowskian
spacetime. In cluster of galaxies for example there exist large scale magnetic fields with $10^{-7}G$ to $10^{-6}G$ on $10 kpc$. More recently Barrow et al \cite{10} have shown that in a general relativistic in  electrodynamic background, can induce cosmic magnetic fields that falls within the interval $10^{-20}G$ to $10^{-12}G$ which are able to seed the galactic dynamo mechanism which amplifies for example in the strongest magnetic field seed, $10^{6}$ orders of magnitude to obtain the microGauss galactic magnetic fields.  Recently we have shown that cosmological magnetic fields
can be obtained from dilatons in torsioned spacetime \cite{3}, where axionic terms such as $F\tilde{F}$ where
${\tilde{F_{\mu\nu}}}={\epsilon}_{\mu\nu\alpha\beta}F^{\alpha\beta}$ are absent. Actually in the case of previous paper we
have a dilaton field in the term $f^{2}(\phi)F^{2}$. However
formally in that previous paper we also use the torsion vector as a
gradient of the axion field like in paper by Mielke and Romero
\cite{11} which however does not address problems of dynamos and
cosmological magnetic fields. In the present paper we use DKO axionic electrodynamics to obtain a seed field of $B_{seed}\sim{10^{-12}Gauss}$ which as we have seen above can feed galactic dynamos.
\section{Dilatonic electromagnetism with torsion and bounds on cosmic magnetism} 
From the lagrangean of dilatonic electromagnetism 
\begin{equation}
{\cal{L}}\sim{{\partial}_{\mu}{\phi}{\partial}^{\mu}{\phi}+f^{2}({\phi})F_{\mu\nu}F^{{\mu}{\nu}}} \label{1}
\end{equation}
the corrected formula is 
\begin{equation}
\dot{f}=\dot{{\phi}}{f¹}
\label{2}
\end{equation}
where the index over f means derivative w.r.t $\phi$. Thus the field equation becomes
\begin{equation}
{\dot{\phi}}^{2}=f^{2}B^{2}
\label{3}
\end{equation}
which yields
\begin{equation}
B\sim{f^{-1}T_{0}}
\label{4}
\end{equation}
where $T_{0}$ here is the actual terrestrial torsion
given in Laemmerzahl \cite{12} given by $T\sim{10^{-17}cm^{-1}}$ or $10^{-31}GeV$. Here the phenomenological $f\sim{10^{-6}GeV^{-1}}$ and since $1G\sim{10^{-20}GeV^{2}}$ which yields
\begin{equation}
{B}\sim{10^{-5}G} \label{5}
\end{equation}
which is a limit found recently by Yokoyama \cite{12}taking into account the virtual pairs and Schwinger effect. If one now take into account another value for torsion for radiogalaxies \cite{13} $T\sim{10^{-46}GeV}$ one obtains a cosmic magnetic field $10^{-10}G$ which is intergalactic magnetic field, thus as far as these results are concerned no dynamo amplification is needed as in Bamba result. New perspectives for the future await a possible finalization of the BICEP 3 undergoing experiment and its final data. 
\section{Magnetogenesis in axionic electrodynamics with torsion}
Recently DKO have obtained an axionic electrodynamics in Riemann-Cartan spacetime with torsion where torsion vector shall be constrained by the gradient of the axion field which strongly suppress some effects of the DKO axionic electrodynamics. potential ${\Psi}$
where torsion vector is given by the gradient of the torsion
potential as $\textbf{T}={\nabla}{\psi}$. DKO theory has a double degree of freedom of torsion from the dynamical point of view, and torsion can be transmutted to an axion. Actually QED can transmute torsion into an axion. Without digging into the effects for string theory we simple start from the DKO axionic electrodynamics throughout the equation of the coupling of axions to photons
\begin{equation}
{\nabla}_{\mu}F^{{\mu}{\nu}}=2{\lambda}{\nabla}_{\mu}{\phi}{\tilde{F}}^{{\mu}{\nu}}
\label{6}
\end{equation}
where ${\nabla}$ is the Riemann-Cartan covariant derivative. Here $\lambda= \frac{e^{2}}{f_{\phi}4{\pi}^{2}}$. Now let us drop the idea of curved spacetime and imagine that our geometrical background is only given by a Minkowski spacetime endowed with torsion. Thus the covariant derivative operator becomes
\begin{equation}
{\nabla}_{\mu}={\partial}_{\mu}+T_{\mu}\label{7} 
\end{equation}
Substitution of this covariant expression into the equation {\ref{1}} yields
\begin{equation}
{\partial}_{\mu}F^{{\mu}{\nu}}+T_{\mu}F^{{\mu}{\nu}}=2{\lambda}[{\partial}_{\mu}{\phi}+T_{\mu}{\phi}]{\epsilon}^{\mu\nu\rho\sigma}F_{\rho\sigma}
\label{8}
\end{equation}
By rescaling the axion field as ${\phi}\rightarrow{ln{\tilde{\phi}}}$ and dropping the tilde in $\phi$ one may define the torsion as
\begin{equation}
T_{\mu}=-{\partial}_{\mu}ln{\phi} \label{9}
\end{equation}
Here $T\sim{10^{-17}cm^{-1}}$ is obtained by
C Laemmerzahl \cite{14} from Hughes-Drever experiment. Therefore with this trick we turn the DKO axionic electrodynamics into Maxwell like equations in Minkowski spacetimes endowed with torsion the only difference being that now torsion is still transmutted into axions and interact indirectly with photons. Thus the homogeneous DKO equation becomes
\begin{equation}
{\partial}_{\mu}F^{{\mu}{\nu}}+T_{\mu}F^{{\mu}{\nu}}=0
\label{10}
\end{equation}
Let us now write down these expressions in terms of the 
vector components of electric and magnetic fields
\begin{equation}
{\nabla}\times\textbf{B}+\textbf{T}\times{\textbf{B}}=0
\label{11}
\end{equation}
\begin{equation}
{\nabla}.\textbf{E}+\textbf{T}.\textbf{E}=0
\label{12}
\end{equation}
Dimensionally the last equation reads
\begin{equation}
{T}\sim{L^{-1}}\label{13}
\end{equation}
where L is the coherence length of the magnetic field. From equation (\ref{11}) one obtains the solution
\begin{equation}
{B_{G}}=B_{seed}e^{TL} \label{14}
\end{equation} 
where $B_{G}$ is the galactic dynamo fixed by observations as $10^{-6}G$ and since torsion is very $10kpc$ the seed magnetic field able to seed galactic dynamo is given by $10^{-12}G$ from torsion contribution which coincides with the highest cosmic magnetic field bound. The future prospects involve solving the complete set of DKO equations in full detail in the non-homogeneous case which includes the axion coupling as well. \section{Conclusions}
The issue of origin of cosmic magnetic field has been a matter of  controversy specially in terms of torsion. Recently K Bamba et al \cite{7} has shown that by using a special type of gravity theory of torsion earlier called by Einstein by the name teleparallelism. one could obtain interstellar magnetic fields without the use of dynamo mechanism. In this paper we show that weaker magnetic fields than the interestellar ones can do be generated by dynamo mechanism which generates the galactic magnetic fields observed in our galaxies, this was possible fro DKO axionic electrodynamics \cite{16}with torsion.

\section{Acknowledgements} We
would like to express my gratitude to A Brandenburg and Prof C Sivaram for helpful discussions on the problem of dynamos
and torsion. Special thanks go to Professor J Yokohama for his many enlightning discussions on the problem of magnetogenesis and to J Jain for discussions on dynamo efficiency both at Institute d'Astrophysique at Paris meeting on Primordial Universe after Planck mission held in december 2014. Financial support from CNPq. and University of State of
Rio de Janeiro (UERJ) are grateful acknowledged.

\end{document}